\documentclass[preprint,12pt]{elsarticle}
\usepackage{graphicx}
\usepackage{dcolumn}
\usepackage{amsmath}
\def\Journal#1#2#3#4{{#1} {#2} (#4) #3 }

\def\NPA{{\rm Nucl. Phys.} A}
\def\NPB{{\rm Nucl. Phys.} B}
\def\PLB{{\rm Phys. Lett.}  B}
\def\PRL{\rm Phys. Rev. Lett.}
\def\PRD{{\rm Phys. Rev.} D}
\def\PRC{{\rm Phys. Rev.} C}

\def\JPG{{\rm J. Phys.} G}

\def\ep{\epsilon}

\def\la{\langle}
\def\ra{\rangle}

\def\be{\begin{equation}}
\def\ee{\end{equation}}
\def\bea{\begin{eqnarray}}
\def\eea{\end{eqnarray}}

\journal{Physics Letters B}

\begin{document}
\begin{frontmatter}
\title{Light-front zero-mode contribution to the
tensor form factors for the exclusive rare $P\to V\ell^+\ell^-$ decays}
\author{Ho-Meoyng Choi}
\ead{homyoung@knu.ac.kr}
\address{ Department of Physics, Teachers College, Kyungpook National University,
     Daegu 702-701, Republic of Korea}

\author{Chueng-Ryong Ji}
\ead{crji@ncsu.edu}
\address{ Department of Physics, North Carolina State University,
Raleigh, NC 27695-8202, USA}

\begin{abstract}
We study the light-front zero-mode contribution to the tensor
form factors $T_i(i=1,2,3)$ for the exclusive rare $P\to V\ell^+\ell^-$ decays
using a covariant fermion field theory model in $(3+1)$ dimensions.
While the zero-mode contribution in principle depends on the form of the
vector meson vertex $\Gamma^\mu=\gamma^\mu - (2k-P_V)^\mu/D$, the three
tensor form factors $T_i(i=1,2,3)$ are found to be free from the zero mode
if the denominator $D$ contains
the term proportional to the light-front energy or the longitudinal momentum
fraction factor $(1/x)^n$ of the struck quark with the power $n>0$.
Since the denominator $D$ used in the light-front quark model (LFQM) has the
power $n=1/2$, the three tensor form factors $T_i(i=1,2,3)$ can be computed in
LFQM safely without involving any complicate zero-mode contribution.
The lack of zero-mode contribution benefits the phenomenology with LFQM.
\end{abstract}

\end{frontmatter}

The study of $B$ meson physics is important not only in extracting the most
accurate values of the Cabibbo-Kobayashi-Maskawa (CKM) matrix elements~\cite{CKM} but also
in searching for new physics effects beyond the standard model (SM). Especially,
the flavor-changing neutral current (FCNC) processes of $B\to K^{(*)}\ell^+\ell^-(\ell=e,\mu,\tau)$
that proceed via loop diagrams~\cite{SM}
in the SM are very important for not only testing the SM but also probing new physics such as
supersymmetric heavy particles in SUSY models, appearing virtually in the loop diagrams to
interfere with those in the SM. While the experimental tests of exclusive decays are
much easier than those of inclusive ones, the theoretical understanding of exclusive
decays is complicated mainly due to the nonperturbative hadronic form factors entered in the
long-distance nonperturbative contributions. Therefore, a reliable estimate of the hadronic
form factors for the exclusive rare $B$ decays is very important for making correct predictions
within and beyond the SM.

Perhaps, one of the most well-suited formulations for the analysis of exclusive processes
involving hadrons may be provided in the framework of light-front (LF) quantization~\cite{BPP}.
For its simplicity and the predictive power of the hadronic form factors
in low-lying ground-state hadrons, especially mesons, the LF
constituent quark model (LFQM) based on the LF quantization has become a very useful and popular
phenomenological tool to study various electroweak properties of
mesons~\cite{Te,Dz,CCP,CC,Ja90,MF97,CJ1,CCH}.
However, the zero-mode complication~\cite{Zero} in the matrix
element has been noticed for the electroweak form factors involving a spin-0 and
spin-1 particles~\cite{Ja99,BCJ02,CJ04}.
The zero mode can be interpreted as residues
of virtual pair creation processes in the $q^+(=q^0+q^3)\to 0$ limit,
i.e., the nonzero contribution from the nonvalence part in
the $q^+=0$ frame~\cite{Zero}. Therefore, finding the zero-mode contribution
correctly in various electroweak transitions is a very important issue in LF
hadron phenomenology.

In our previous works, we have studied the zero-mode contribution to the
hadronic form factors for $P\to P$~\cite{Bc09} and $P\to V$~\cite{ZM05,ZM10} transitions,
where $P$ and $V$ stand for pseudoscalar and vector mesons, respectively. Using an
exactly solvable covariant Bethe-Salpeter (BS) model~\cite{BCJ02}, we have developed
the method that correctly pin down the existence/absence of the
zero-mode contribution to the form factors. Our method of finding the zero-mode
contribution is based on a direct power counting~\cite{BCJ02} of
the longitudinal momentum fraction $x(0\leq x \leq 1)$ in the $q^+\to 0$ limit for the off-diagonal
elements in the Fock-state expansion of the current matrix.
Since the longitudinal momentum
fraction is one of the integration variables in the LF matrix elements (i.e. helicity
amplitudes), our power counting method is straightforward as far as we know the
behaviors of the longitudinal momentum fraction in the integrand.
In the analysis of the $P\to P(V)$ transitions,
we used the phenomenologically accessible pseudoscalar vertex
$\Gamma^\mu_{\rm P}=\gamma^5$ and vector meson vertex $\Gamma^\mu_{\rm V}=\gamma^\mu - (P_V-2k)/D$,
where $k$ and $P_V-k$ are the relative four-momenta for the constituent quark and
 antiquark, respectively, and $P_V$ is the four-momentum of the vector meson.
For the manifestly covariant model, we used two different cases of the denominator $D$ for the
vector meson vertex,
i.e.
(1) $D=D_{\rm cov}(M_V)=M_V + m_q + m_{\bar q}\sim (1/x)^0$, and
(2) $D=D_{\rm cov}(k\cdot P_V)=[2k\cdot P_V + M_V (m_q + m_{\bar q})-i\epsilon]/M_V\sim (1/x)^1$,
where $m_q$ ($m_{\bar q}$) is the constituent quark (antiquark) mass and
$M_V$ is the physical vector meson mass.
We also discussed the application of the LF version of the denominator
$D_{\rm LF}(M_0)=M_0 + m_q + m_{\bar q}\sim (1/x)^{1/2}$ with the
invariant mass $M_0$ of the vector meson,  which has been widely used in
the LFQM phenomenology~\cite{Ja90,CCH,Ja99,C07,HW07,QM10}.
For the exclusive $P\to P$ decays, we analyzed both semileptonic $P\to P\ell\nu_\ell$ and
rare $P\to P\ell^+\ell^-$ decays~\cite{Bc09}.
In the analysis of the hadronic form factors ($f_\pm, f_T$) for the rare $P\to P$ decay~\cite{Bc09},
we found that the form factors $f_+$ and $f_T$ are immune to the zero mode but
the form factor $f_-$ receives the zero mode.
For the exclusive $P\to V$ decays, we analyzed
the semileptonic $P\to V\ell\nu_\ell$ decays~\cite{ZM05,ZM10}.
In the analysis of the weak form factors ($g, a_\pm, f$) for the semileptonic $P\to V\ell\nu_\ell$
decays, we found that the form factors ($g, a_+, f$) are immune to the zero mode but the
form factor $a_-$ receives the zero mode. We also identified the zero-mode
operator for the form factors $f_-$~\cite{Bc09} and $a_-$~\cite{ZM10} that is convoluted with the
initial- and final-state LF wave functions.

The purpose of this Letter is to extend our LF covariant analysis to include
the rare $P\to V\ell^+\ell^-$ decays, where the three tensor form factors $T_i(i=1,2,3)$ are
 necessary for the calculation of the amplitude
in addition to the weak form factors ($g, a_\pm, f$).

The tensor form factors
$\la J^\mu_h\ra_0\equiv\la V(P_2,\ep^*_h)|\bar{q}i\sigma^{\mu\nu}q_\nu b|P(P_1)\ra$
and
$\la J^\mu_h\ra_5\equiv\la V(P_2,\ep^*_h)|\bar{s}i\sigma^{\mu\nu}q_\nu\gamma_5 b|P(P_1)\ra$
for the rare $P\to V\ell^+\ell^-$ decays are
defined~\cite{AW} as
\bea
\la J^\mu_h\ra_0
&=& i \varepsilon^{\mu\nu\alpha\beta}\ep^*_{\nu}P_\alpha q_\beta T_1(q^2),
\nonumber\\
\la J^\mu_h\ra_5
&=& [\epsilon^{*\mu}(P\cdot q)-(\epsilon^*\cdot q)P^\mu]T_2(q^2)
+(\epsilon^*\cdot q)\biggl[ q^\mu -\frac{q^2}{(P\cdot q)}P^\mu\biggr] T_3(q^2),
\nonumber\\
{\label{eq:1p}}
\eea
where $P=P_1+P_2$ and $q=P_1-P_2$ is the four-momentum
transfer to the lepton pair and $4m^{2}_{\ell}\leq q^{2}\leq(M_1-M_2)^2$.
The polarization vector $\epsilon^*_h$ of the final-state vector
meson satisfies the Lorentz condition $\epsilon^*_h \cdot P_2 = 0$.
We also use the convention $\sigma^{\mu\nu}=(i/2)[\gamma^\mu,\gamma^\nu]$ for
the antisymmetric tensor.

\begin{figure}
\begin{center}
\includegraphics[width=3.5in,height=1.2in]{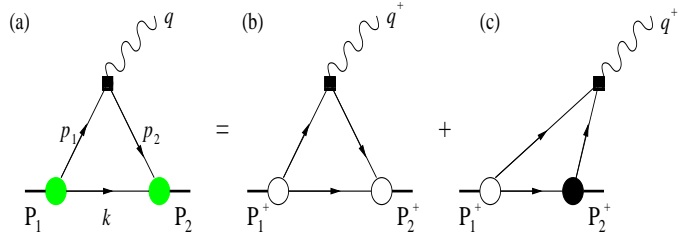}
\end{center}
\caption{ The covariant diagram (a) corresponds to the sum of the
LF valence diagram (b) defined in $0<k^+<P^+_2$ region and the
nonvalence diagram (c) defined in $P^+_2<k^+<P^+_1$ region. The large white
and black blobs at the meson-quark vertices in (b) and (c) represent
the ordinary LF wave functions and the non-wave-function vertex,
respectively. The small black box at the quark-gauge boson vertex
indicates the insertion of the relevant Wilson operator. \label{fig1} }
\end{figure}

The solvable model, based on the covariant BS model of $(3+1)$-dimensional
fermion field theory~\cite{MF97,BCJ02},
enables us to derive the transition form factors between
pseudoscalar and vector mesons explicitly.
The covariant diagram shown in Fig.~\ref{fig1}(a) is in general equivalent to the sum of the LF
valence diagram [Fig.~\ref{fig1}(b)] and the nonvalence diagram [Fig.~\ref{fig1}(c)].
The matrix element
$\la J^\mu_h\ra_{0(5)}$ obtained from the
covariant diagram of Fig. 1(a) is given by
\be\label{eq:4}
\la J^\mu_h\ra_{0(5)}
= i g_1 g_2\Lambda^2_1\Lambda^2_2
\int\frac{d^4k}{(2\pi)^4} \frac{(S^\mu_h)_{0(5)}} {N_{\Lambda_1}
N_{1} N_{\bar q} N_{2} N_{\Lambda_2}},
\ee
where $g_1$ and $g_2$ are the normalization factors  which can be fixed by
requiring charge form factors of pseudoscalar and vector mesons to be
unity at $q^2=0$, respectively. To regularize the covariant fermion
triangle loop in $(3+1)$ dimensions, we replace the
point gauge boson vertex $\gamma^\mu (1-\gamma_5)$ by a non-local
(smeared) gauge boson vertex
$({\Lambda_1}^2 / N_{\Lambda_1})\gamma^\mu (1-\gamma_5)
( {\Lambda_2}^2 / N_{\Lambda_2})$, where $N_{\Lambda_1}
=p_1^2-{\Lambda_1}^2+i\ep$ and $N_{\Lambda_2}
=p_2^2-{\Lambda_2}^2+i\ep$, and  $\Lambda_1$
and $\Lambda_2$ play the role of momentum cut-offs similar to the Pauli-Villars
regularization.  The rest of the
denominators in Eq.~(\ref{eq:4}) coming from the intermediate fermion
propagators in the triangle loop diagram are given by
$N_1 = p_1^2 -{m_1}^2 + i\ep$,
$N_{\bar q} = k^2 - m^2 + i\ep$, and
$N_{2} = p_2^2 -{m_2}^2 + i\ep$,
where $m_1$, $m$, and  $m_2$ are the masses of the constituents
carrying the intermediate four-momenta $p_1=P_1 -k$, $k$, and $p_2=P_2
-k$, respectively.

The trace terms $(S^\mu_h)_{0}$ and $(S^\mu_h)_{5}$ are given by
\bea
(S^\mu_h)_0 &=& {\rm Tr}[(\not\!p_2 + m_2)i\sigma^{\mu\nu}q_\nu
(\not\!p_1 +m_1)\gamma_5(-\not\!k + m)\epsilon^*_h\cdot\Gamma],
 \nonumber\\
(S^\mu_h)_5 &=& {\rm Tr}[(\not\!p_2 + m_2)i\sigma^{\mu\nu}q_\nu\gamma_5
(\not\!p_1 +m_1)\gamma_5(-\not\!k + m)\epsilon^*_h\cdot\Gamma],
{\label{eqq:4}}
\eea

where the final-state vector meson vertex operator $\Gamma^\mu$
is given by
\be\label{eq5}
\Gamma^\mu =\gamma^\mu-\frac{(P_2-2k)^\mu}{D}.
\ee
In this work, we shall analyze with the $D$ factor,
$D_{\rm cov}(M_V)=M_2 + m_2 + m$, for the explicit comparison between the manifestly covariant
calculation and the LF one. However, since the more realistic $D$ factor used in most popular LF
quark model is either $D_{\rm cov}(k\cdot P_2)
=[2k\cdot P_2 + M_2(m_2 + m) -i\epsilon]/M_2$ or $D_{\rm LF}(M'_0)=M'_0 + m_2 + m$ with the invariant mass
$M'_0$ for the final vector meson state, we also discuss the results for both
$D_{\rm cov}(k\cdot P_2)$ and $D_{\rm LF}(M'_0)$ cases.

In the manifestly covariant calculation of Fig.~\ref{fig1}(a), we decompose the product of five denominators
given in Eq.~(\ref{eq:4}) into a sum of terms containing
three propagators. We then use the Feynman parametrization for the three propagators
and make a Wick rotation of Eq.~(\ref{eq:4}) in $d=4-2\epsilon$ dimensions to
regularize the integral, since otherwise one looses the logarithmically
divergent terms in Eq.~(\ref{eq:4}). Following the above procedure (see~\cite{Bc09,ZM10} for more
details), one can easily
obtain the manifestly covariant form factors  $T_i(q^2)(i=1,2,3)$.

Performing the LF calculation of Figs.~\ref{fig1}(b) and~\ref{fig1}(c)
in parallel with the manifestly covariant calculation,
we first choose $q^+ >0$ frame and then take $q^+\to 0$ limit to check the existence/absence of
the zero-mode contribution to the hadronic matrix element given by Eq.~(\ref{eq:4}).
We also use the plus component of the currents to obtain the tensor form factors.
While the form factor $T_1(q^2)$ can be obtained from  $\la J^+_h\ra_0$ with $h=1$,
the form factors $T_2(q^2)$ and $T_3(q^2)$ can be obtained from  $\la J^+_h\ra_5$
with both $h=0$ and $1$.
In the reference frame where $q^+ > 0$ and ${\bf P}_{1\perp}=0$, the
(timelike) momentum transfer $q^2$ is given by
\begin{equation}
{\label{tq2}}
q^2=q^+q^- - {\bf q}^2_\perp =\Delta\biggl(M^2_1 -
\frac{M^2_2}{1-\Delta}\biggr) -\frac{{\bf q}^2_\perp}{1-\Delta},
\end{equation}
where $\Delta =q^+/P^+_1$.
In this frame, the covariant diagram Fig.~\ref{fig1}(a) corresponds to the sum of the
LF valence diagram (b) defined in $0<k^+<P^+_2$ region and the
nonvalence diagram (c) defined in $P^+_2<k^+<P^+_1$ region. The large white
and black blobs at the meson-quark vertices in (b) and (c) represent
the ordinary LF wave functions and the non-wave-function vertex,
respectively. The small black box at the quark-gauge boson vertex
indicates the insertion of the relevant Wilson operator.
We should note that in the $q^+\to 0$ (i.e. $\Delta\to 0$) limit, the nonvalence
region (i.e. $0<x<\Delta$) of integration shrinks to zero. Thus, if the integrand
has a singularity in $p^-_1\sim 1/x$, the nonvalence region may give a nonvanishing
zero mode in the $q^+\to 0$ limit.
In the LF calculations for the trace terms $(S^\mu_h)_{0(5)}$
in Eq.~(\ref{eqq:4}), we separate the on-mass-shell propagating
part $S^\mu_{\rm on}$ from
the off-mass-shell instantaneous part $S^\mu_{\rm inst}$, i.e.
$S^\mu_h = (S^\mu_h)^{\rm on} + (S^\mu_h)^{\rm inst}$ via
$\not\!p + m =(\not\!p_{\rm on} + m) + \frac{1}{2}\gamma^+(p^- - p^-_{\rm on})$.
While the on-mass-shell propagating part $(S^\mu_h)^{\rm on}$ indicates that all
three quarks are on their respective mass-shell, i.e. $k^-=k^-_{\rm on}$
and $p^-_{i}=p^-_{i\rm on}(i=1,2)$, the instantaneous part
$(S^\mu_h)^{\rm inst}$ includes the term proportional
to $\delta p_i^-=p^-_i-p_{i\rm on}^-$ and
$\delta k^-=k^- - k^-_{\rm on}$.
The relations between the current matrix
elements and the form factors in the $q^+=0$ [or Drell-Yan(DY)] frame~\cite{DYW} are as follows:
\bea{\label{ch:4}}
\la J^+_{h=1}\ra_0&=&2\sqrt{2}\varepsilon_{+-12}P^+_1 q^L T_1^{\rm DY}(q^2),
\nonumber\\
 \la J^+_{h=1}\ra_5&=& -\sqrt{2}P^+_1 q^L
 \biggl[ T^{\rm DY}_2 + \frac{q^2}{P\cdot q} T^{\rm DY}_3 \biggr],
\nonumber\\
\la J^+_{h=0}\ra_5 &=& \frac{q^2}{M_2}P^+_1
\biggl [T^{\rm DY}_2 - \frac{ P\cdot q - q^2}{P\cdot q} T^{\rm DY}_3 \biggr],
\eea
where $q^{R(L)}=q_x \pm iq_y$.

In the valence region $0<k^+<P^+_2$ (i.e. $\Delta<x<1$), the pole $k^-=k^-_{\rm on}=({\bf
k}^2_\perp + m^2_{\bar q} -i\ep)/k^+$ (i.e., the spectator quark) is located in the lower half of
the complex $k^-$-plane.  Thus, the Cauchy integration formula for the
$k^-$ integral in Eq.~(\ref{eq:4}) gives
  \be\label{va1}
 \la J^+_h\ra_{0(5)} =
\frac{N}{16\pi^3}\int^1_0 \frac{dx}{(1-x)}\int d^2{\bf k}_\perp
\chi_1(x,{\bf k}_\perp)(S^+_h)_{0(5)}\chi_2 (x, {\bf k'}_\perp),
  \ee
where $N=g_1g_2\Lambda^2_1\Lambda^2_2$ and
$(S^+_h)_{0(5)}=(S^+_h)^{\rm on}_{0(5)}+(S^+_h)^{\rm inst}_{0(5)}$.
The LF vertex functions $\chi_1$ and $\chi_2$ are given by
 \be\label{va2}
 \chi_{1(2)}(x,{\bf k}^{(\prime)}_\perp) =
 \frac{1}{ x^2 (M^2_{1(2)} - M^{(\prime)2}_0)(M^2_{1(2)}-M^{(\prime)2}_{\Lambda_{1(2)}})},
\ee
where ${\bf k'}_\perp={\bf k}_\perp + (1-x) {\bf q}_\perp$ and
$M^{(')2}_0 = ({\bf k}^{(\prime)2}_\perp + m^2)/(1-x)
+ ({\bf k}^{(\prime)2}_\perp + m^2_{1(2)})/x$ and
$M^{(\prime)2}_{\Lambda_{1(2)}}= M^{(\prime)2}_0(m_{1(2)}\to\Lambda_{1(2)})$.
We note that only the on-mass-shell propagating parts $(S^+_h)^{\rm on}_{0(5)}$
contribute to the valence region in Eq.~(\ref{va1}).
The valence contributions to the form factors
$T_i(q^2)(i=1,2,3)$ in the $q^+=0$ frame are obtained as

\bea\label{JPval}
T_1^{\rm DY}(q^2) &=& \frac{N}{8\pi^3}\int^1_0
\frac{dx}{(1-x)^2}\int d^2{\bf k}_\perp\; \chi_1\chi_2
\biggl\{ (2x-1){\bf k}^2_\perp
+ (1-x){\bf k}_\perp\cdot{\bf q}_\perp
\nonumber\\
&&+ {\cal A}_1 {\cal A}_2
-2(1-x)\frac{({\bf k}_\perp\cdot{\bf q}_\perp)^2}{q^2}
+ 2(1-x)\frac{(m_1+m_2)}{D}
\nonumber\\
&&\times\biggl[ {\bf k}^2_\perp
+ \frac{({\bf k}_\perp\cdot{\bf q}_\perp)^2}{q^2} \biggr]
\biggr\},
\eea
\bea\label{JPval2}
T^{\rm DY}_2(q^2) &=&\frac{N}{8\pi^3}
\int^1_0
\frac{dx}{(1-x)^2}\int d^2{\bf k}_\perp\; \chi_1\chi_2
\biggl\{ {\bf k}^2_\perp
+ (2x-1)(1-x){\bf k}_\perp\cdot{\bf q}_\perp
\nonumber\\
&&+ {\cal A}_1 {\cal A}_2
+ 2(1-x) \frac{({\bf k}_\perp\cdot{\bf q}_\perp)^2}{{\bf q}^2_\perp}
- \frac{2(1-x)}{D}[ (1-x)q^2 - {\bf k}_\perp\cdot{\bf q}_\perp]
\nonumber\\
&&\times\biggl[{\cal A}_1  - (m_1 + m_2)\frac{{\bf k}_\perp\cdot{\bf q}_\perp}{q^2}\biggr]
\biggr\} - \frac{q^2}{P\cdot q}T^{\rm DY}_3(q^2),
\eea
\bea\label{JPval3}
T^{\rm DY}_3(q^2) &=&\frac{N}{8\pi^3}
\int^1_0\frac{dx}{(1-x)^2}\int d^2{\bf k}_\perp\; \chi_1\chi_2
\biggl\{ {\bf k}^2_\perp + {\cal A}_1 (2m - {\cal A}_2)
\nonumber\\
&&+ (1-x){\bf k}_\perp\cdot{\bf q}_\perp - 2(1-x)\frac{({\bf k}_\perp\cdot{\bf q}_\perp)^2}{q^2}
- 2\frac{{\bf k}_\perp\cdot{\bf q}_\perp}{q^2}
 [{\bf k}^2_\perp
 \nonumber\\
 &&+ x(1-x)M^2_2 + xm^2 + (1-x)(m_1m + m_2m - m_1m_2)]
\nonumber\\
&&-\frac{2}{D}[{\bf k}^2_\perp + m^2 - (1-x)^2M^2_2  + (1-x){\bf k}_\perp\cdot{\bf q}_\perp]
\nonumber\\
&&\times\biggl[{\cal A}_1 - (m_1 + m_2)\frac{{\bf k}_\perp\cdot{\bf q}_\perp}{q^2}\biggr]
\biggr\},
\eea

where ${\cal A}_i = (1-x) m_i + xm(i=1,2)$. When we consider only the simple vector meson vertex
$\Gamma^\mu_V=\gamma^\mu$ (i.e. $1/D=0$),
our LF results $T^{\rm DY}_i(q^2)(i=1,2,3)$ obtained from the valence contributions in
the $q^+=0$ frame are exactly the same as the
manifestly covariant results. That is, there are no zero-mode contributions to the form factors $T_i(q^2)(i=1,2,3)$. Indeed, our LF results $T^{\rm DY}_i(q^2)(i=1,2,3)$ are also immune
to the zero mode even if we include the more realistic $D$ factor such as $D_{\rm cov}(k\cdot P_2)$ and
$D_{\rm LF}(M'_0)$. Only if we use the naive $D$ factor such as $D_{\rm cov}(M_V)=M_V + m_2 + m$,
the zero-mode contribution exists in the matrix element of $\la J^+_{h=0}\ra_5$.

For the completeness of the analysis, we shall identify the zero-mode contribution
to $\la J^+_{h=0}\ra_5$ for the $D=D_{\rm cov}(M_V)$ case.
In the nonvalence region $P^+_2<k^+<P^+_1$ (i.e. $0<x<\Delta$), the poles are at $p^-_1=p^-_{1\rm
on}(m_1) = [m^2_1 +{\bf k}^2_\perp -i\ep]/p^+_1$ (from the struck quark propagator)
and $p^-_1=p^-_{1\rm on}(\Lambda_1) = [\Lambda^2_1+{\bf k}^2_\perp -i\ep]/
p^+_1$ (from the smeared quark-gauge-boson vertex), which are located in the upper
half of the complex $k^-$-plane.
For $D=D_{\rm cov}(M_V)$ case, we find the suspected zero-mode terms,
i.e. singular terms proportional to $p^-_1$ in the off-mass-shell propagator
$(S^+_{h=0})^{\rm inst}_5$ as follows
 \be\label{ZMS}
 (S^+_{h=0})^{\rm Z.M.}_5=\lim_{\Delta\to 0}(S^+_{h=0})^{\rm inst}_5
= \frac{4p^-_1 }{M_2 D}[ m_1 {\bf q}^2_\perp
  - (m_1 + m_2){\bf p}_{1\perp}\cdot{\bf q}_\perp].
 \ee
We note that the suspected zero-mode terms $(S^+_{h=0})^{\rm Z.M.}_5$ in Eq.~(\ref{ZMS})
leads to the nonvanishing zero-mode contribution to $\la J^+_{h=0}\ra_5$ when
$D=D_{\rm cov}(M_V)$ due to the singular behavior $p^-_1/D_{\rm cov}(M_V)\sim 1/x$.
Following the similar procedure discussed in~\cite{Bc09,ZM10},
we can identify the zero-mode operator that is convoluted with
the initial- and final-state valence wave functions to generate
the zero-mode contribution. Explicitly, the zero-mode contribution $\la J^+_{h=0}\ra^{\rm Z.M.}_5$ or
$T_3^{\rm Z.M.}=(M_2/q^2)\la J^+_{h=0}\ra^{\rm Z.M.}_5$ can
be expressed in terms of the zero-mode operator convoluted with the initial- and final-state LF
vertex functions:
\bea\label{eq:j9}
T_3^{\rm Z.M.}(q^2)&=&\frac{N}{8\pi^3}\frac{2}{D_{\rm cov}(M_V)}
\int^1_0\frac{dx}{(1-x)}\int d^2{\bf k}_\perp
\chi_1 \chi_2
\nonumber\\
&&\times\biggl\{(m_1+m_2)\biggl[ A^{(1)}_2Z_2
+ \frac{q\cdot P}{q^2}A^{(2)}_1\biggr]
 - m_1 Z_2\biggr\},
 \eea
 where $A^{(1)}_2$, $A^{(2)}_1$, and $Z_2$ are given by~\cite{Ja99,Bc09}
 \bea\label{Z2}
 &&A^{(1)}_2 = \frac{x}{2} + \frac{{\bf k}_\perp\cdot{\bf q}_\perp}{q^2},\;
 A^{(2)}_1 = -{\bf k}^2_\perp - \frac{({\bf k}_\perp\cdot{\bf q}_\perp)^2}{q^2},
 \nonumber\\
 &&Z_2 = x(M^2_1 - M^2_0) + m^2_1 - m^2 + (1-2x)M^2_1
- [q^2 + q\cdot P]\frac{{\bf k_\perp\cdot{\bf q}_\perp}}{q^2}.
\nonumber\\
 \eea
 By adding
$T_3^{\rm Z.M.}(q^2)$ to $T_3^{\rm DY}(q^2)$ in Eqs.~(\ref{JPval2}) and ~(\ref{JPval3}),
i.e. $T_3^{\rm Full}(q^2)=T_3^{\rm DY}(q^2) + T_3^{\rm Z.M.}(q^2)$,
we confirm that our LF
results for the form factors $T_2$ and $T_3$ are in an exact agreement with the manifestly covariant
results for the $D=D_{\rm cov}(M_V)$ case.

However, $T_3^{\rm Z.M.}(q^2)$ vanishes when $D=D_{\rm cov}(k\cdot P_2)$
or $D_{\rm LF}(M'_0)$ is used. This can be easily seen from
the power counting rule for $x$ (or $p^-_1$) in $x\to 0$ limit.
Note that $D_{\rm cov}(k\cdot P_2)\sim x^{-1}$ and $D_{\rm LF}(M'_0)\sim x^{-1/2}$
while $p^-_1\sim x^{-1}$ in the same $x\to 0$ limit.
Since $p^-_1/D_{\rm cov}(k\cdot P_2)\sim x^{0}$ and $p^-_1/D_{\rm LF}(M'_0)\sim x^{-1/2}$,
the case of $D_{\rm cov}(k\cdot P_2)$ or $D_{\rm LF}(M'_0)$ provides less singular behavior
than the case of $D_{\rm cov}(M_V)$.
More detailed power counting rule
for the longitudinal momentum fraction can be found in~\cite{Bc09,ZM05,ZM10}.
Therefore, as far as the $D_{\rm LF}(M'_0)$ or $D_{\rm cov}(k\cdot P_2)$ is used in
 LFQM, there are no zero-mode contributions to the tensor form factors $T_i(q^2)(i=1,2,3)$
 and the form factors given by Eqs.~(\ref{JPval})-(\ref{JPval3}) are the correct LF tensor
 form factors.

\begin{figure}
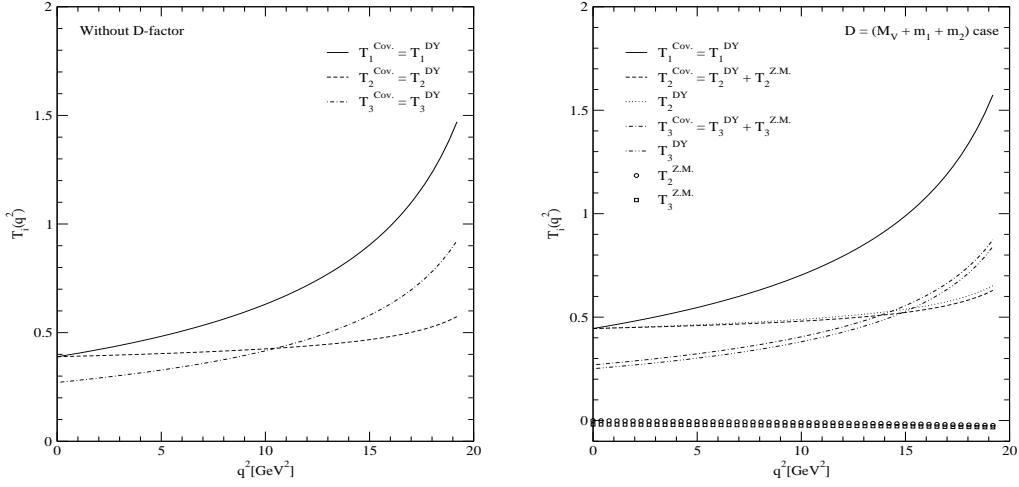

\includegraphics[width=2.5in, height=2.5in]{Fig2.eps}
\hspace{0.5cm}
\includegraphics[width=2.5in, height=2.5in]{Fig3.eps}
\caption{The tensor form factors $T_i(q^2)(i=1,2,3)$ for
$B\to K^*$ transition for the simple vector meson vertex
$\Gamma^\mu=\gamma^\mu$ case (i.e. without the $D$ factor) and
for $\Gamma^\mu=\gamma^\mu-(P_2-2k)^\mu/D$ with the $D=D_{\rm cov}(M_V)$,
respectively.}
\label{fig2}
\end{figure}

In Fig.~\ref{fig2}, we show both the manifestly covariant and the LF results of the
tensor form factors $T_i(q^2)(i=1,2,3)$ for the rare
$B\to K^*\ell^+\ell^-$ transition obtained from the exactly solvable covariant BS
model of fermion field theory. The used model parameters for $B$ and $K^*$ mesons are
$m_b=4.9$ GeV, $m_s=0.5$ GeV, $m_{u(d)}=0.43$ GeV, $\Lambda_1=10$ GeV,
$\Lambda_{2}=2.5$ GeV, $g_1=5.13$, and $g_2=3.2$.
The left and right panels of Fig.~\ref{fig2} are the results
for the simple vector meson vertex
$\Gamma^\mu=\gamma^\mu$ case (i.e. $1/D=0$) and
for $\Gamma^\mu=\gamma^\mu-(P_2-2k)^\mu/D$ with $D=D_{\rm cov}(M_V)$,
respectively. Our LF results $T^{\rm DY}_i(q^2)(i=1,2,3)$ obtained from the $q^+=0$
frame are analytically continued to the timelike $q^2 > 0$ region by
changing ${\bf q}^2_\perp$ to $-q^2$ in the form factor. The results for the more
realistic covariant vertex with $D=D_{\rm cov}(k\cdot P_2)$ are basically the same
as those for the simple vertex case although the quantitative behaviors are slightly
different from each other,
i.e., the LF tensor form factors
$T_i^{\rm DY}(q^2)$ given by Eqs.~(\ref{JPval})-(\ref{JPval2}) with $D=D_{\rm cov}(k\cdot P_2)$
are exact results without encountering the zero-mode contribution.
For the $D=D_{\rm LF}(M'_0)$ case, although we do not know how to
compute the nonvalence diagram,
we can still use our counting rule for the longitudinal momentum
fraction factors, i.e. $p^-_1/D_{\rm LF}(M'_0)\to(1/x)^{1/2}$ as
$\Delta\to 0$, to check the existence of the zero mode. As we discussed,
the zero-mode contribution from $p^-_1/D_{\rm LF}(M'_0)$
does not exist as in the case of $D= D_{\rm cov}(k\cdot P_2)$.

In Table~\ref{t1}, we summarize our findings of the existence/absence of the zero-mode
contribution to the hadronic form factors ($g, a_\pm, f$)~\cite{ZM05,ZM10} and $T_i(i=1,2,3)$
for the rare $P\to V\ell^+\ell^-$ decays depending on the components of the weak currents $J^\mu_{V-A}$, polarization vector $\epsilon^*_h$ of the vector meson, and various $D$ factors in $\Gamma^\mu_{\rm V}$.
Since our findings of the existence/absence of the zero mode are based on the method of power counting, our conclusion applies to other methods of regularization as far as the regularization doesn't change the power counting in the form factor calculation. For example, as discussed by Jaus in Ref.~\cite{Ja99}, some other multipole type ansatz in the method of regularization wouldn't change the conclusion drawn by our monopole type ansatz.
This may exemplify the benefit
of our method to remove any unnecessary caution regarding on the possible
zero-mode contribution in the hadron phenomenology by correctly pinning
down its existence/absence.

\begin{table}
\caption{The existence ($\rm O$) or absence ($\rm X$) of the zero-mode contribution to the
form factors for the rare $P\to V\ell^+\ell^-$ decays depending on the components of the
currents $J^\mu$, polarization vector
$\epsilon^*_h$ of the vector meson, and various $D$ factors in $\Gamma^\mu_{\rm V}$.}
\label{t1}
\renewcommand{\tabcolsep}{0.3pc} 
\begin{tabular}{@{}lccccccc} \hline
 & $g$ & $a_+$ & $a_-$ & $f$ & $T_1$ & $T_2$ & $T_3$ \\
\hline
$\la J^\mu_h\ra$ & $\la J^+_{1}\ra_V$ & $\la J^+_{1}\ra_A$
& $\la J^+_{0,1}\ra_A$,$\la J^\perp_{1}\ra_A$ & $\la J^+_{0,1}\ra_A$
& $\la J^+_1\ra_0$ & $\la J^+_{0,1}\ra_5$ & $\la J^+_{0,1}\ra_5$\\
\hline
$\Gamma^\mu_{\rm V}=\gamma^\mu$ & $\rm X$ & $\rm X$ & $\rm O$ & $\rm X$ & $\rm X$ & $\rm X$ & $\rm X$ \\
$\Gamma^\mu_{\rm V}[D_{\rm cov}(M_V)]$ & $\rm X$ & $\rm X$ & $\rm O$ & $\rm O$ & $\rm X$ & $\rm O$ & $\rm O$ \\
$\Gamma^\mu_{\rm V}[D_{\rm cov}(k\cdot P_2)]$ & $\rm X$ & $\rm X$ & $\rm O$ & $\rm X$ & $\rm X$ & $\rm X$ & $\rm X$ \\
$\Gamma^\mu_{\rm V}[D_{\rm LF}(M'_0)]$ & $\rm X$ & $\rm X$ & $\rm O$ & $\rm X$ & $\rm X$ & $\rm X$ & $\rm X$ \\
\hline
\end{tabular}
\end{table}

In this work, we have analyzed the zero-mode contribution to the tensor
form factors for the rare $P\to V\ell^+\ell^-$ decays.
For the phenomenologically accessible vector
meson vertex $\Gamma^\mu_{\rm V}=\gamma^\mu - (P_2-k)^\mu/D$, we discussed the three
typical cases of the $D$ factor which also may be classified by the differences
in the power counting of the LF energy (or longitudinal momentum fraction $x$)
$p^-_1\sim 1/x$, i.e.:
(1) $D_{\rm cov}(M_V)=M_V + m_2 + m\sim (1/x)^0$,
(2) $D_{\rm cov}(k\cdot P_2)=[2k\cdot P_2 + M_2(m_2 + m) -i\epsilon]/M_2 \sim (1/x)^1$,
and
(3) $D_{\rm LF}(M'_0)=M'_0 + m_2 + m\sim (1/x)^{1/2}$.
Our main idea to obtain the transition form factors is first to find if the
zero-mode contribution exists or not for the given form factor using the power counting
method. If it exists, then the separation of the on-mass-shell propagating part from
the off-mass-shell part is useful since the latter is responsible for the
zero-mode contribution. We found that the form factor $T_1(q^2)$ is immune
to the zero-mode contribution in all three cases of the $D$ factors.
However, for the form factors $T_2(q^2)$ and $T_3(q^2)$, while the zero-mode contribution exists in the
$D_{\rm cov}(M_V)$ case, the other two cases such as $D_{\rm cov}(k\cdot P_2)$
and $D_{\rm LF}(M'_0)$ are immune to the zero-mode contribution.
We also should note that the zero-mode contribution does not exist in the simple vector meson
vertex $\Gamma^\mu=\gamma^\mu$ (i.e. $1/D=0$ case).

All of these findings stem from the fact that the zero-mode contribution from the $D$ factor
is absent if the denominator $D$ of the vector meson vertex
$\Gamma^\mu=\gamma^\mu - (P_2-k)^\mu/D$
contains the term proportional to the LF energy (or longitudinal momentum fraction $x$)
$(p^-_1)^n\sim (1/x)^n$ with the power $n>0$. Since the phenomenologically accessible LFQM
satisfies this condition $n>0$, only the valence contributions to the three weak form factors
($g, a_+, f$) and the three tensor form factors $T_i(q^2)(i=1,2,3)$ obtained in the
$q^+=0$ frame for the analysis of the rare $P\to V\ell^+\ell^-$ decays
are sufficient to provide the full results of the LFQM. Although the form factor
$a_-(q^2)$ receives the zero-mode contribution, it comes only from the simple vertex $\gamma^\mu$
term but not from the $D$ factor satisfying the condition $n>0$. In this case, we were able to
obtain the correct zero-mode operator that is convoluted with the initial- and final-state
LF wave functions~\cite{ZM10}.
Our analyses of zero-mode contribution can at least assure
the Lorentz invariance of the hadron form factors in the exclusive
processes that we have considered up to now.
 This certainly benefits the hadron phenomenology and the application to
the more realistic LFQM analysis for various semileptonic and rare $P\to V$ decays is underway.

\section*{Acknowledgment}
 The work of H.-M.Choi was supported by the Korea
Research Foundation Grant funded by the Korean
Government(KRF-2010-0009019) and that of C.-R.Ji by the U.S.
Department of Energy(No. DE-FG02-03ER41260).

\end{document}